\documentclass[
10pt,tightenlines, amsmath, amssymb, nofootinbib,
showpacs, preprintnumbers]{revtex4}

\newcommand{\beq}{\begin{equation}}
\newcommand{\eeq}{\end{equation}}
\newcommand{\bea}{\begin{eqnarray}}
\newcommand{\eea}{\end{eqnarray}}
\newcommand{\bmat}{\begin{pmatrix}}
\newcommand{\emat}{\end{pmatrix}}

\newcommand{\junk}[1]{}

\def\pb{\overline{\psi}}
\def\gm{\gamma^{\mu}}
\def\d{\partial}
\def\<{\langle}
\def\>{\rangle}
\def\+{\dagger}
\def\D{{\cal D}}
\def\Dd{{\cal D^{\dagger}}}
\def\Ht{H^{\bot}}

\def\hlv{\frac{1}{2}}

\begin{document}
\title{Currents on Superconducting Strings at Finite Chemical
Potential and Temperature} \author{Max~A.~Metlitski}
\email{mmetlits@physics.ubc.ca} \affiliation{Department of Physics
and Astronomy, University of British Columbia, Vancouver, BC,
Canada, V6T 1Z1}
\date{\today}
\begin{abstract}
We consider the model giving rise to Witten's superconducting
cosmic strings at finite fermion chemical potential and
temperature. We demonstrate how various symmetries of the
hamiltonian can be used to exactly compute the fermion electric
current in the string background. We show that the current along
the string is not sensitive to the profiles of the string fields,
and at fixed chemical potential and temperature depends only on
the string winding number, the total gauge flux through the vortex
and, possibly, the fermion mass at infinity.
\end{abstract}
\maketitle
\section{Introduction}
Ever since Witten's pioneering paper\cite{Witten}, it has been
known that cosmic strings can posses fermion zero modes
concentrated in the string core. One remarkable feature of this
system, is that an application of a constant electric field in the
string direction induces an electric current along the string
carried by the zero modes. This current will grow linearly with
time, while the electric field is turned on and will persist even
after the field is turned off. The string, thus, becomes
superconducting. It must be noted, however, that the behaviour of
the system is known precisely only for induced currents smaller
than a certain critical current - once the current exceeds this
critical value, the energies of the zero modes become larger than
the fermion mass at infinity $m$, and it becomes possible for the
charge carriers to move off the string, quenching the
superconductivity. The question of build up of charge and current
on the superconducting string in an external electric field has
been analyzed extensively in \cite{Witten, Harvey, Naculich,
GordonSC, Widrow, Forte}.

In this paper, we investigate a very different mechanism for
inducing a current on the string. Namely, we compute the current
on a superconducting string in the presence of a non-zero fermion
chemical potential $\mu$ and temperature $T$. It is a rather
trivial exercise to calculate the current $J$ along the string for
$\mu \lesssim m$, $T \ll m$, when then the low energy dynamics of
the fermion-string system are governed by an effective $1+1$
dimensional theory of zero modes moving along the string. In this
case it is straightforward to show that the current for each
fermion species is $J = \frac{e \mu n}{2 \pi}$, where $e$ is the
fermion charge and $n$ is the winding number of the string.
However, we make a much stronger statement: it is possible to
calculate exactly the total electric current in the string
direction for $\it{any}$ value of the fermion chemical potential
$\mu$ and temperature $T$. We shall show that the result is
topological in nature, and is independent of the particular
profiles of the background string fields. The result will depend
crucially on whether the string is local (as considered by Witten)
or global (as, for instance, in the case of axion strings). In
particular, if the string is local, the naive prediction for $J$
of the effective $1+1$ dimensional theory, remains valid for any
value of $\mu$, $T$.

The appearance of quantum numbers (particularly of fermion number)
on topological defects is a very well-developed subject with known
computational methods\cite{GordonRev}, such as trace identities
and adiabatic expansion. At zero chemical potential, the fermion
charge induced on defects is usually a topological quantity and,
frequently, can be evaluated exactly. However, at arbitrary finite
chemical potential, the fermion charge induced is generally not
topological\cite{Niemi}, and difficult to compute exactly. In view
of this, our result is particularly interesting, since we show
that quantum numbers such as total current can remain topological
and exactly calculable at arbitrary fermion chemical potential.
Mathematically, our analysis can be easily generalized to a large
class of Hamiltonians involving fermions in $d+1$ dimensions in
the background of a $d$ dimensional defect, which is uniform in
the $(d+1)$'st direction.

As far as we know, the problem of computation of electric current
on a superconducting string in the background of an arbitrary
fermion chemical potential and temperature has not been considered
before, although some of the techniques we use have been
previously discussed in conjunction to index
theorem\cite{Weinberg} for string zero modes, and charge induced
on the string by an electric field\cite{GordonSC, Forte}. Although
the present problem could be of some interest in application to
cosmology, 
this paper was largely motivated by related problems in dense
quark matter. It is well known that quark matter at large baryon
chemical potential, which might be realized in the cores of
neutron stars, breaks certain symmetries of $QCD$\cite{CFLRev},
and may supports several kinds of strings\cite{ForbesZ}. Recently,
the method of fictitious axial anomalies has been used\cite{SZ} to
derive the following effective action for the interaction of an
electromagnetic field $A_{\mu}$ with an axial string in dense
quark matter: \beq S = \sum_a\frac{e_a \mu_a n}{2 \pi}
\frac{Q_a}{q} \int A_i dl^i\eeq Here the line integral is along
the string, the index $a$ runs over all species of quarks, and the
fraction $\frac{Q_a}{q}$ denotes the flavor content of the
condensate that supports the axial string. This implies that the
axial string in dense $QCD$ carries an electric current of $J =
\frac{e_a \mu_a n}{2 \pi} \frac{Q_a}{q}$ for each quark species.
We wish to understand this phenomenological result
microscopically. The method for computation of current on strings
at finite chemical potential developed in\cite{SZ} is sensitive
only to the pattern of symmetry breaking and, thus, would yield in
application to the present model the same result, $J = \frac{e \mu
n}{2 \pi}$. As we see below, in some cases this result remains
correct for all $\mu$, while in other cases it receives
corrections of order $1$.

\section{Currents on Strings}
\subsection{The Model} Consider the following model of a Dirac fermion $\psi$
coupled to a string: \beq  \label{L} {\cal L} = \pb i \gm
(\d_{\mu} - i e A_{\mu} - \frac{i q}{2} R_{\mu} \gamma^5)  \psi -
h \pb (\phi^{*} \frac{1+\gamma^5}{2} + \phi \frac{1-\gamma^5}{2})
\psi \eeq Here $A_{\mu}$ and $R_{\mu}$ are gauge fields and $\phi$
is a complex scalar field.  The model has the following classical
gauge symmetries: \bea U(1)&:& \psi \rightarrow e^{i e \alpha(x)}
\psi, \;
A_{\mu} \rightarrow A_{\mu} +  \d_{\mu} \alpha, \; \phi \rightarrow \phi \\
\tilde{U}(1)&:& \psi \rightarrow e^{i q \theta(x) \gamma^5/2}
\psi, \; R_{\mu} \rightarrow R_{\mu} + \d_{\mu} \theta, \; \phi
\rightarrow e^{i q \theta(x)} \phi \eea This model is exactly
equivalent to Witten's model of superconducting cosmic strings
with a particular choice of gauge charges\footnote{We could have
easily considered the completely general version of Witten's
model, however, to simplify the algebra slightly we concentrate on
the above choice of gauge charges, which makes the $A_{\mu}$ field
couple to the vector current and the $R_{\mu}$ field couple to the
axial current.}. By convention, we associate the vector field
$A_{\mu}$ with electromagnetism. We note that the above model
suffers from gauge anomalies, which can be removed, for example,
by adding another fermion $\hat{\psi}$ to the model with the
opposite $R$ charge $\hat{q} = -q$ and the electric charge
$\hat{e}$, such that ${\hat{e}}^2 = e^2$. The Lagrangian for the
$\hat{\psi}$ fermion is then: \beq \label{Lh} {\cal \hat{L}} =
\hat{\pb} i \gm (\d_{\mu} - i \hat{e} A_{\mu} + \frac{i q}{2}
R_{\mu} \gamma^5) \hat{\psi} - h \hat{\pb} (\phi
\frac{1+\gamma^5}{2} + \phi^{*} \frac{1-\gamma^5}{2}) \hat{{\psi}}
\eeq Notice that $\hat{\psi}$ now couples to $\phi^{*}$ rather
than $\phi$.

We could also consider the situation when the $\tilde{U}(1)$
symmetry is global, such that the gauge field $R_{\mu}$ is absent,
the Lagrangian (\ref{L}) is by itself anomaly free, and the
addition of the $\hat{\psi}$ field is unnecessary. In our
calculations, we will recover this case by taking $q = 0$.

We assume that the $\tilde{U}(1)$ symmetry is spontaneously
broken, the $\phi$ field acquires a non-zero expectation value and
 strings of the $\phi$ field are possible. We wish to consider
the fermion $\psi$ in the background of an infinitely long static
string uniform in the $z$ direction. The string is characterized
by a non-zero winding
 number $n$ of the scalar
field: \beq n = \int \frac{dl^a}{2 \pi i} \frac{\phi^{*} \d_a
\phi}{|\phi|^2}\eeq where the integral is over a contour in the
$xy$ plane at infinity, and the absolute value of the scalar field
$|\phi|$ tends to some constant $\phi_0$ as $r \rightarrow \infty$
in the $xy$ plane.

If the $\tilde{U}(1)$ symmetry is local, then in most models (such
as the Abelian Higgs model), the condition that the string energy
is finite, implies that $D_{\mu}\phi = (\d_{\mu} - i q R_{\mu})
\phi  \rightarrow 0$ fast enough as $r \rightarrow \infty$ in the
$xy$ plane. This in turn implies the quantization of the string
flux: \beq \label{Flux} \Phi = \frac{q}{4 \pi} \int d^2 x
\epsilon^{ab} R_{ab} = n\eeq From here on $a,b = 1,2$ and
$R_{\mu\nu}$ is the usual field strength tensor. It must be noted
that the condition (\ref{Flux}) is not present in the case of
global strings, so we will throughout our calculations keep the
flux $\Phi$ arbitrary and at the end set $\Phi = n$ for local
strings
and $\Phi = 0$ for global strings. 

Our objective is to calculate the expectation value of the
electromagnetic fermion current in the string direction, \beq J^3
= e\int d^2 x \langle \pb \gamma^3 \psi\rangle \eeq at finite
fermion chemical potential $\mu$ and temperature $T$. Note that if
the $\tilde{U}(1)$ symmetry is local, there is an additional
contribution to the electromagnetic current from the $\hat{\psi}$
fermions. This, however, can be obtained from the result for the
$\psi$ fermions by setting $e \rightarrow \hat{e}$, $q \rightarrow
\hat{q} = -q$, $\phi \rightarrow \phi^{*}$, which translates
into $\Phi \rightarrow -\Phi$, $n \rightarrow -n$. 
\subsection{The Spectrum}
Let's start by analyzing the spectrum of our fermions in the
string background. The one-particle Hamiltonian is: \beq \label{H}
H = -i \alpha^i (\d_i - i e A_i - \frac{iq}{2} R_i \gamma^5) + h
\gamma^0 (\phi^{*} \frac{1+\gamma^5}{2} + \phi
\frac{1-\gamma^5}{2}) \eeq where $\alpha^i = \gamma^0 \gamma^i$
and $i = 1,2,3$. For a static background string uniform in the
third direction, $A_i = 0$, $R_3 = 0$, and hence, \bea H &=& -i
\d_3
\alpha^3 + \Ht\\
\Ht &=& -i \alpha^a (\d_a - \frac{i q}{2} R_a \gamma^5) + h
\gamma^0 (\phi^{*} \frac{1+\gamma^5}{2} + \phi
\frac{1-\gamma^5}{2}) \eea Since all the fields are assumed
uniform in the third direction, we can choose $-i \d_3 \psi = k
\psi$ and work at fixed $k$.\footnote{We take the third direction
$z$ to be compact of length $L$ so that the eigenvalues $k$ are
discrete. As usual, we will take $L \rightarrow \infty$ at the end
of the calculation.} In each $k$ sector, \beq H_k = k \alpha^3 +
\Ht\eeq and the operator $H_k$ now acts solely in the transverse
$xy$ plane. At this point, we make our choice of the $\gamma$
matrices to be: \beq \alpha^3 = \left(\begin{array}{cc} 1 & 0\\0 &
-1 \end{array}\right), \;\; \alpha^a = \left(\begin{array}{cc} 0 & i \sigma^a\\
-i \sigma^a & 0 \end{array}\right), \;\; \gamma^0 = \left(\begin{array}{cc} 0 & 1 \\
1 & 0 \end{array}\right), \;\; \gamma^5 = \left(\begin{array}{cc} \sigma^3 & 0 \\
0 & -\sigma^3 \end{array}\right) \eeq The operator $\Ht$ then
takes the form: \bea \Ht &=& \left(\begin{array}{cc} 0 & \D \\
\Dd & 0 \end{array}\right)\eea where, \bea\D &=& \d_a \sigma^a +
\frac{q}{2} R_a \epsilon^{ab} \sigma^b + h(\frac{1+\sigma^3}{2}
\phi + \frac{1-\sigma^3}{2}\phi^{*}) \eea

Let's discuss the properties of the operator $\Ht$. Since
$|\phi|\rightarrow \phi_0$ as $r\rightarrow \infty$, the continuum
spectrum of $\Ht$ starts at eigenvalues $|\lambda| = m = |h|
\phi_0$. $\Ht$ may also have bound states. We let $m_b$ be the
smallest positive eigenvalue of $\Ht$. By dimensional reasons,
$m_b \sim m$. Now, observe, \beq \{\alpha^3, \Ht\} = 0 \eeq Thus,
$\alpha^3$ maps a properly normalized eigenstate $|\lambda\>$ of
$\Ht$ with eigenvalue $\lambda$ into a properly normalized
eigenstate of $\Ht$ with eigenvalue $-\lambda$. Moreover, since
$\alpha^3$ maps zero modes of $\Ht$ into zero modes of $\Ht$, all
the zero-modes of $\Ht$ can be classified by their eigenvalue
under $\alpha^3$. Writing, $\lambda(x) = (u(x), \, v(x))$, we note
that the zero modes of $\Ht$ with $\alpha^3 = 1$ satisfy $v = 0$,
$\Dd u = 0$, while the zero modes of $\Ht$ with $\alpha^3 = -1$
satisfy $u = 0$, $\D v = 0$. So letting $N_+$ be the number of
$\alpha^3 = 1$ zero modes, and $N_-$ the number of $\alpha^3 = -1$
zero modes, we have, \beq N = N_+ - N_- = dim(ker(\Dd)) -
dim(ker(\D)) = Index(\Ht)\eeq Hence, $N$ is the index of an
elliptic operator, which is usually a strongly topological
quantity. $N$ has been first computed explicitly for a particular
background string configuration in \cite{JackiwR} to be: \beq N =
n\eeq This result was later generalized
\cite{GordonSC},\cite{Weinberg} to arbitrary background string
fields.

We now return to the operator $H_k$. Observe, $[H_k, {\Ht}^2] =
0$. So, we can obtain the spectrum of $H_k$ from the spectrum of
$\Ht$ in the following way. Let, \beq \Ht \lambda(x) = \lambda\,
\lambda(x) \eeq First, suppose, $\lambda
> 0$. Then, the state, \beq \label{corr} \psi(x) = c_1 \lambda(x) + c_2
\alpha^3 \lambda(x) \eeq is going to be an eigenstate of $H_k$
with eigenvalue $E$, provided that, \beq \left(\begin{array}{cc} \lambda & k \\
k & -\lambda \end{array}\right) \left(\begin{array}{c} c_1 \\
c_2 \end{array}\right) = E \left(\begin{array}{c} c_1 \\
c_2 \end{array}\right)\eeq The eigenvalues of the above equation
are, \beq E = \pm \sqrt{\lambda^2 + k^2} \eeq and the
eigenvectors, \beq\label{eigenvecs} {\left(\begin{array}{c} c_1 \\
c_2 \end{array}\right)}_{\pm} = \frac{1}{{(2{(\lambda^2 +
k^2)}^{\hlv})}^{\hlv}}\left(\begin{array}{c} \pm sgn(k)
{({(\lambda^2 + k^2)}^{\hlv}\pm \lambda)}^{\hlv} \\ {({(\lambda^2
+ k^2)}^{\hlv}\mp \lambda)}^{\hlv}
 \end{array}\right)\eeq Thus, each eigenstate of $\Ht$ with positive
eigenvalue, generates one positive energy and one negative energy
eigenstate of $H_k$. However, this correspondence has to be taken
with a grain of salt, since most eigenstates of $\Ht$ are
continuum states, and the ``1 to 2" map discussed above between
eigenstates of $\Ht$ and eigenstates of $H_k$ need not preserve
the density of states.

The zero modes of $\Ht$ are also simultaneously eigenstates of
$H_k$. These have the dispersion, \beq E = k \alpha^3 \eeq So the
zero modes of $\Ht$ become chiral fermions moving up or down the
string depending on the sign of their eigenvalue under $\alpha^3$.

\subsection{Current - Naive Approach}
We now proceed to the computation of electric current at finite
$\mu, T$. This is given by:\beq \label{trace}J^3 = e\int d^2x
\,\<\pb \gamma^3 \psi\rangle = e\int d^2 x \, tr\langle x|\alpha^3
n(H) |x\rangle \eeq where, \beq n(E) =
\frac{sgn(E)}{e^{\beta(E-\mu)sgn(E)} + 1} \eeq is the usual
Fermi-Dirac distribution. Summing over each momentum sector $k$,
we obtain\footnote{Here $tr$ denotes matrix trace and $Tr$ denotes
a general operator trace.}, \beq \label{trTr}J^3 = e\frac{1}{L}
\sum_k \int d^2 x\, tr\langle x|\alpha^3 n(H_k)|x\rangle  = e
\frac{1}{L} \sum_k Tr(\alpha^3 n(H_k))\eeq Using the
correspondence between spectra of $H_k$ and $\Ht$, we may {\it
schematically} write the operator trace (\ref{trTr}) as: \beq
\label{bigtrace}J^3 = e \frac{1}{L} \sum_k \sum_{E(H_k)} \langle
\psi_E| \alpha^3| \psi_E \rangle n(E) = e \frac{1}{L} \sum_k
\left( \sum_{\lambda(\Ht) > 0,\,s = \pm} \langle
\psi_{\lambda,k,s}| \alpha^3| \psi_{\lambda,k,s}\rangle
n(E_{s}(\lambda,k))  + \sum_{\lambda(\Ht) = 0} \langle
\lambda|\alpha^3|\lambda\rangle n(E(\lambda,k))\right) \eeq Here
$E(H_k)$ denote eigenstates of $H_k$, $\lambda(\Ht)$ denote
eigenstates of $\Ht$, and $\psi_{\lambda, k, \pm}$ denote
eigenstates of $H_k$ generated by an eigenstate $|\lambda\rangle$
of $\Ht$, with energies $E_{\pm}(\lambda,k) = \pm \sqrt{\lambda^2
+ k^2}$. Again, we stress that the above representation would have
been absolutely correct if all the states contributing to the
operator trace were discreet, and normalizable (for instance if $T
= 0$ and $\mu < m$). In our case, this is not generally so, but we
choose for now to ignore this problem, in order to illustrate the
general idea behind the computation. We will later return to take
the continuum states into consideration more carefully.

For the moment suppose, $T = 0$, $0 < \mu < m_b$. Then $n(E) =
\theta(E) \theta (\mu - E)$. Hence, only states generated by zero
modes contribute to the sum in (\ref{bigtrace}), as all the other
states have energies $|E| = \sqrt{\lambda^2 + k^2} \geq |\lambda|
\geq m_b
> \mu$. The zero modes are eigenstates of $\alpha^3$, and thus,
satisfy, $E = \alpha^3 k$ and $\langle \lambda |\alpha^3 |\lambda
\rangle = \alpha^3$. Thus, \beq J^3 = e \frac{1}{L}\sum_{k}(N_+
\theta(\mu - k) \theta(k) - N_-\theta(\mu +k) \theta(-k)) = e
N\int \frac{dk}{2 \pi} \,\theta(k) \theta(\mu - k) = \frac{e \mu
N}{2 \pi} = \frac{e \mu n}{2 \pi} \eeq where we've used the fact
that index $N$ is equal to the winding number of the vortex $n$.

Now, let's relax our assumption and work at arbitrary $T, \mu$. We
first need to evaluate the matrix element $\langle
\psi_{\lambda,k,s}| \alpha^3| \psi_{\lambda,k,s}\rangle$ (in what
follows we suppress the indices $\lambda, k, s$). Using
eq.(\ref{corr}) and $({\alpha^3})^2 = 1$, we see $\langle \psi|
\alpha^3| \psi\rangle = (|c_1|^2 + |c_2|^2)\langle
\lambda|\alpha^3|\lambda\rangle + (c_1^* c_2 + c_2^* c_1)$.
Recalling $\alpha^3|\lambda\rangle = |-\lambda\rangle$, we obtain,
$\langle \psi|\alpha^3|\psi\rangle = c_1^* c_2 + c_2^* c_1 =
\frac{k}{E}$, where we've used eq. (\ref{eigenvecs}). Hence,
\beq\label{tracel2} J^3 = e \frac{1}{L} \sum_k \left(
\sum_{\lambda(\Ht)
> 0,\,s = \pm} \frac{k}{E_{s}(\lambda,k)} n(E_{s}(\lambda,k))  +
N_+ n(k) - N_- n(-k)\right)\eeq Now, observe that
$E_{s}(\lambda,k) = E_{s}(\lambda,-k)$ for $\lambda > 0$. Hence,
the first sum in the brackets in eq. (\ref{tracel2}) is odd in
$k$, and, thus, the contribution to $J^3$ from non-zero modes of
$\Ht$ cancels out exactly, leading to: \beq \label{Jnaive1}J^3 = e
N \frac{1}{L} \sum_k n(k) = e n \int \frac{dk}{2\pi} n(k) = e n
\,n_0(\mu, T)\eeq Here $n_0(\mu,T)$ is the number density of a
free massless chiral fermion in $1$ dimension, at finite chemical
potential $\mu$ and temperature $T$. It is a peculiar fact that
$n_0(\mu,T)$ is temperature independent and equals
$\frac{\mu}{2\pi}$, so that, \beq \label{Jnaive2} J^3 = \frac{e
\mu n}{2 \pi} \eeq 

\subsection{Current - Corrections from Polarized Continuum} Although, the
result (\ref{Jnaive2}) is very attractive, it is actually
generally {\it incorrect}. We know that this result is exact for
$T = 0$, $\mu < m$, when $J^3$ receives contributions only from
normalizable eigenstates of $H_k$. We will now show, that the
presence of long range vortex fields polarizes the continuum
eigenstates of $H_k$ in a way, which might significantly modify
the result (\ref{Jnaive2}) for $\mu > m$.

Let's return to the trace (\ref{trTr}). We can rewrite this
expression in terms of spectral current density as: \bea
\label{J3int}J^3 &=& \int dE \,n(E) j^3(E)\\ \label{j3E} j^3(E)
&=& e \frac{1}{L} \sum_k Tr(\alpha^3 \delta(H_k - E)) \eea We use
the following representation of the delta function, $\delta(x)
=\frac{i}{2\pi} \lim_{\epsilon\rightarrow 0^+}
\left(\frac{1}{x+i\epsilon} -\frac{1}{x-i\epsilon}\right)$, to
rewrite, \beq \label{lim} j^3(E) = \frac{i}{2
\pi}\lim_{\epsilon\rightarrow 0^+} e \frac{1}{L} \sum_k
Tr\left(\alpha^3 (\frac{1}{H_k + z^+} - \frac{1}{H_k +
z^-})\right) \eeq where $z^+ = -E + i \epsilon$, $z^- = -E -
i\epsilon$. From here on, the limit $\epsilon \rightarrow 0^+$ is
implied. Simplifying (\ref{lim}), \beq \label{J31} j^3(E) =
\frac{1}{L} \sum_k\frac{i e}{2 \pi} Tr\left(\alpha^3 (\frac{H_k -
z^+}{H_k^2 - (z^+)^2} - \frac{H_k - z^-}{H_k^2 - (z^-)^2})\right)
=\frac{1}{L} \sum_k \frac{i e}{2 \pi} Tr\left(\alpha^3 (\frac{k
\alpha^3 + \Ht - z^+}{{\Ht}^2 + k^2 - (z^+)^2} - \frac{k \alpha^3
+ \Ht - z^-}{{\Ht}^2 + k^2 - (z^-)^2})\right) \eeq where we've
used $H_k^2 = {\Ht}^2 + k^2$. The terms in (\ref{J31}), which are
odd in $k$ cancel out, and we obtain, \beq \label{J32} j^3(E) =
\frac{1}{L} \sum_k \frac{i e}{2 \pi} Tr\left(\alpha^3 (\frac{\Ht -
z^+}{{\Ht}^2 + k^2 - (z^+)^2} - \frac{\Ht - z^-}{{\Ht}^2 + k^2 -
(z^-)^2})\right) \eeq Now, $\{\alpha^3, \Ht\} = 0$. Hence, for any
function $f$, $Tr(\alpha^3 \Ht f({\Ht}^2)) = -Tr (\Ht
f({\Ht}^2)\alpha^3) = - Tr (\alpha^3 \Ht f({\Ht}^2)) = 0$, and,
\beq \label{J33} j^3(E) = \frac{1}{L} \sum_k \frac{i e}{2 \pi}
Tr\left(\alpha^3 (\frac{- z^+}{{\Ht}^2 + k^2 - (z^+)^2} - \frac{-
z^-}{{\Ht}^2 + k^2 - (z^-)^2})\right)\eeq We now introduce the
function $g$, \beq g(M^2) = Tr\left(\alpha^3 \frac{M^2}{{\Ht}^2 +
M^2}\right) = Tr\left(\frac{M^2}{\D\Dd + M^2}\right) -
Tr\left(\frac{M^2}{\Dd \D + M^2}\right) \eeq This function is very
well known \cite{GordonRev},\cite{Weinberg} as it satisfies, \beq
N = Index(\Ht) = \lim_{M^2 \rightarrow 0} g(M^2) \eeq More
generally, $g(M^2)$ is related to the spectral asymmetry
$\sigma_k(E)$, of the Hamiltonian $H_k$\cite{GordonRev}, and hence
to its $\eta$-invariant as, \bea \label{sigma} \sigma_k(E) &=&
\frac{i}{2 \pi} k \left(G(k^2 - (z^+)^2) - G(k^2 -
(z^-)^2)\right)\\
\eta_k &=& 2\int_{0}^{\infty} dE \; \sigma_k(E) \\ G(z) &=&
\frac{g(z)}{z}\eea Here, $g(z)$ is understood as the analytic
continuation of $g$ from $\Re_{+}$ to ${\cal C}$. From eq.
(\ref{J33}), we can express $j^3(E)$ in terms of $G$ as,
\beq\label{J34} j^3(E) = \frac{1}{L} \sum_k \frac{-i e}{2
\pi}\left(
 z^+ G(k^2 - (z^+)^2) - z^-G(k^2 - (z^-)^2)\right) = \frac{1}{L}\sum_k e \, \frac{E}{k}\, \sigma_k(E)
 \eeq
 Following the technique of trace identities
described in detail by \cite{GordonRev},\cite{Weinberg}, one can
explicitly calculate $g(M^2)$ to be: \beq g(M^2) = n - (n-\Phi)
\frac{M^2}{m^2 + M^2}\eeq Hence, the index $N = \lim_{M^2
\rightarrow 0} g(M^2) = n$, in agreement with previous
calculations \cite{GordonSC},\cite{Weinberg},\cite{JackiwR}.
Continuing $g$ to the complex plane, we obtain, \beq \label{G}
G(z) = \frac{n}{z} - (n-\Phi)\frac{1}{z+m^2} \eeq Hence,
generically, G has a pole at $z = 0$, and a pole at $z = -m^2$,
i.e. at the continuum threshold. Notice, however, that the pole at
$z = m^2$ disappears when $n = \Phi$.

We can now substitute the result (\ref{G}) into (\ref{sigma}) and
take the limit $\epsilon \rightarrow 0^+$ to calculate the
spectral asymmetry, \beq \label{sigmar}\sigma_k(E) = k \, sgn(E)
(n \, \delta(E^2 - k^2) - (n-\Phi) \delta(E^2 - k^2 - m^2)) \eeq
which yields the $\eta$-invariant, \beq \label{eta} \eta_k = n\,
sgn(k) - (n-\Phi) \frac{k}{\sqrt{k^2 + m^2}} \eeq We note that eq.
(\ref{eta}) is in agreement with previous calculation of the
$\eta$-invariant\cite{GordonSC}.

Returning to the evaluation of current, we substitute the result
(\ref{sigmar}) into eq. (\ref{J34}) to obtain, \beq j^3(E) =
\frac{1}{L} \sum_k \frac{e}{2} \left(n (\delta(E -k) + \delta(E +
k)) - (n-\Phi) (\delta(E - \sqrt{k^2 + m^2}\,) + \delta(E +
\sqrt{k^2 + m^2}\,))\right)\eeq and the total current in the
string direction (\ref{J3int}) becomes, \beq J^3 = e \int
\frac{dk}{2\pi}  \left(n \, n(k) - \frac{n-\Phi}{2} (n(\sqrt{k^2 +
m^2}) + n(-\sqrt{k^2 + m^2}))\right)\eeq This can be conveniently
rewritten as, \beq \label{J3final} J^3 = e (n\,n_0(\mu,T)
-\frac{n-\Phi}{2}n_m(\mu,T)) \eeq where $n_0(\mu,T) = \frac{\mu}{2
\pi}$ is the familiar number density of one-dimensional chiral
massless fermions, and, \beq n_m(\mu,T) = \int \frac{dk}{2 \pi}
(n(\sqrt{k^2 + m^2}) + n(-\sqrt{k^2 + m^2}))\eeq is the number
density of one-dimensional 2-component (Dirac) fermions of mass
$m$.

Several comments are in order here. First of all, we see from eq.
(\ref{J3final}) that the naive result (\ref{Jnaive1}) is generally
modified by a contribution from modes located at the continuum
threshold. Observe, that for $\mu = 0$, the current $J^3$ vanishes
for all temperatures. At non-zero chemical potential, two cases
are of particular interest. The first case is that of a local
string, satisfying the finite energy condition, $D\phi \rightarrow
0$ faster than $1/r$, which implies $\Phi = n$. In this case, the
contribution from continuum modes vanishes, and we recover our
initial result (\ref{Jnaive2}), which is due solely to the zero
modes, \beq J^3 = \frac{e \mu n}{2 \pi} \eeq This ``coincidence"
can be explained as follows. If $D\phi \rightarrow 0$ fast enough,
the fields in the problem are, essentially, short range, and hence
we can easily put the system in a box, making the spectrum
discrete, so that the argument in section II C is correct.

Let's briefly discuss what happens when we add the second fermion
$\hat{\psi}$ to the problem. Recall, we used this fermion to
cancel gauge anomalies of our model. As noted in section II A, the
contribution of $\hat{\psi}$ to $J^3$ can be obtained by taking $e
\rightarrow \hat{e}$, $\mu \rightarrow \hat{\mu}$, $n \rightarrow
-n$, $\Phi \rightarrow -\Phi$. In particular, the continuum modes
at threshold again cancel out, and $\hat{J^3} = -\frac{\hat{e}
\hat{\mu}n}{2\pi}$. In particular, if the chemical potentials of
$\psi$ and $\hat{\psi}$ fermions are the same, we can obtain a
non-vanishing total electromagnetic current along the string, by
letting\footnote{This choice certainly respects the anomaly
cancellation condition $\hat{e}^2 = e^2$.} $\hat{e} = -e$ , so
that, \beq J^3_{EM} = \frac{e {\mu} n}{\pi} \eeq

The second practically interesting case is that of a global
string. This case can be recovered by taking $\Phi \rightarrow 0$.
Then, \beq J^3 = e n (n_0(\mu,T) - \frac{1}{2} n_m(\mu,T)) \eeq In
this case, the field $\phi$ is long range, and there is a
significant modification of the result (\ref{Jnaive1}). Note that
$n_m(\mu,T)$ is no longer temperature independent, so for
simplicity, we choose to work at $T = 0$, $\mu > 0$. Then, $n(E) =
\theta(E) \theta(\mu - E)$ and, \beq J^3 =\frac{e n}{2 \pi} (\mu -
\theta(\mu - m)\sqrt{\mu^2 - m^2}\,) \eeq Thus, for $\mu < m$, the
current is governed by our original result (\ref{Jnaive2}), while
for $\mu > m$, we also get a counterflow current from the states
at continuum threshold. Thus, $J^3(\mu)$ has a cusp at $\mu = m$,
and for $\mu \gg m$ falls off to $0$ as $\frac{e m n}{4\pi}\,
\frac{m}{\mu}$.

\section{Conclusion}
In this paper, we have found an exact expression for the electric
current on superconducting strings as a function of fermion
chemical potential and temperature. We've analyzed the case of
both local and global strings, and our analysis has not been
limited to a low energy theory of zero modes in the string core.
Our ability to obtain such an exact result has been due to a
cancellation (or partial cancellation) of contributions of all,
but the zero fermion modes to the current. For local strings,
we've seen that for {\it all} values of $T$ and $\mu$, the current
is due to zero modes in the string core. On the other hand, for
global strings, the current receives contributions both from the
zero modes and from certain states at continuum threshold. The
latter contribution tends to cancels out the contribution from the
zero modes as the fermion chemical potential becomes much larger
than the fermion mass $m$. The results for $\mu \gg m$, might be
particularly interesting in application to currents on axial
strings in dense quark matter\cite{SZ}, where the gap $\Delta \ll
\mu$.

We would like to note that the study of persistent topological
currents and spin currents in conjunction with problems, such as,
for example, Quantum Hall Effect\cite{Laughlin} and Spin-Hall
Effect\cite{Zhang}, has over the past years become an active
subject of research in condensed matter physics. It would be
interesting to investigate the relation of the phenomenon
discussed in this paper to problems in condensed matter systems.
For instance, persistent supercurrents, are known to appear on
vortices in superfluid $^3He-A$ and, somewhat similarly to
currents considered in this paper, are due to chiral
anomalies\cite{Volovik}.

\section*{Acknowledgements}
I would like to acknowledge very helpful discussions with
A.~R.~Zhitnitsky, D.~T.~Son, G.~E.~Volovik and P.~B.~Wiegmann. I
would also like to thank the organizers of the program ``QCD and
Dense Matter: From Lattices to Stars" at the Institute for Nuclear
Theory, Seattle, where this work was initiated. This work was
supported in part by the Natural Sciences and Engineering Research
Council of Canada.


\begin{thebibliography}{10}
\bibitem{Witten}E.~Witten, Nucl.\ Phys.\ B {\bf 249} (1985) 557.
\bibitem{Harvey}C.~G.~Callan and J.~A.~Harvey,
Nucl.\ Phys.\ B {\bf 250} (1985) 427.
\bibitem{Naculich}
S.~G.~Naculich,
Nucl.\ Phys.\ B {\bf 296} (1988) 837.
\bibitem{GordonSC}G.~W.~Semenoff, Phys.\ Rev.\ D {\bf 37} (1988) 2838.
\bibitem{Forte}S.~Forte, Phys.\ Rev.\ D {\bf 38} (1988) 1108.
\bibitem{Widrow}L.~M.~Widrow, Phys.\ Rev.\ D {\bf 38} (1988) 1684.
\bibitem{GordonRev}A.~J.~Niemi and G.~W.~Semenoff, Phys.\ Rep.\
{\bf 135} (1986) 99.
\bibitem{Niemi}
A.~J.~Niemi,
Nucl.\ Phys.\ B {\bf 251} (1985) 155.
\bibitem{Weinberg}E.~Weinberg, Phys.\ Rev.\ D {\bf 24} (1981) 2669.
\bibitem{CFLRev}K.~Rajagopal and F.~Wilczek, hep-ph/0011333.
\bibitem{ForbesZ}
M.~M.~Forbes and A.~R.~Zhitnitsky,
Phys.\ Rev.\ D {\bf 65} (2002) 085009, hep-ph/0109173.
\bibitem{SZ} D.~T.~Son and A.~R.~Zhitnitsky, Phys.\ Rev.\ D. {\bf
70} (2004) 074018, hep-ph/0405216.
\bibitem{JackiwR}
R.~Jackiw and P.~Rossi,
Nucl.\ Phys.\ B {\bf 190} (1981) 681.
\bibitem{Laughlin}
R.~B.~Laughlin, Rev.\ Mod.\ Phys.\ {\bf 71} (1999) 863.
\bibitem{Zhang}
S.~Murakami, N.~Nagaosa and S.~C.~Zhang,
Science {\bf 301} (2003) 1348.
\bibitem{Volovik}
M.~M.~Solomaa, G.~E.~Volovik, Rev.\ Mod.\ Phys.\ {\bf 59} (1987)
533.
\end{thebibliography}
\end{document}